# Star Formation Processes versus Planet Formation Processes

Shiv S. Kumar

*The Galileo Institute, P.O. Box 6516, Charlottesville, VA 22906, U.S.A. & Department of Astronomy, University of Virginia, Charlottesville, VA, U.S.A.*

**Abstract.** The processes of star formation are fundamentally different from those of planet formation. Since the mass of a very-low-mass object alone doesn't allow us to uniquely determine its basic nature, we have to look at its other characteristics, such as its motion, its age, its atmospheric composition, its internal structure and composition, etc., in order to ascertain its formation mechanism.

As I have repeatedly pointed out, the processes of star formation are fundamentally different from those of planet formation (Kumar 1964; Kumar 1967; Kumar 1974; Kumar 1990; Kumar 2002). Stars are, in general, formed by the fragmentation of interstellar or primordial clouds whereas planets generally are formed by the slow accumulation (accretion) of dust, rocks, and gas in the vicinity of a star. The stellar domain, which exists independent of the planetary domain, may extend down to mass of ~ 0.001 Msun (or ~ 1 Mjup). In the planetary domain, the maximum mass of an object formed by the planet formation processes (in the vicinity of a star of any mass) is ~ 2 Mjup (Kumar 2002).

In the past few years, quite a few people in the scientific community have referred to the luminous and dark objects with mass below 0.013 Msun (the so-called deuterium burning limit) as 'planets', but that, frankly speaking, is illogical. Just because a gaseous object of mass 0.01 Msun, formed by the star formation processes, doesn't go through the deuterium burning reactions doesn't mean that it's fundamentally different from a gaseous object of mass 0.016 Msun formed by the same formation mechanism. The destruction of deuterium in the interior of a young, contracting, very-low-mass gaseous object doesn't change the structure or final destiny of the object; all it does is to slow down its evolution a bit (Kumar 1963b). Whether or not they go through deuterium burning, all hydrogen-rich, very-low-mass gaseous objects (with mass below the Kumar limit of ~ 0.08 Msun) quickly end up as completely degenerate objects (Kumar 1962; Kumar 1963a; Kumar 1963b).

Since deuterium burning is irrelevant to determining the basic nature of a very-low mass luminous or dark object, we have to look at other characteristics such as its age, its atmospheric chemical composition, its internal structure and composition, its motion, etc. to ascertain its formation mechanism. The determination of the basic nature of an object solely based on its mass is likely to lead to incorrect conclusions. Let us, for example, briefly look at the case of the dark companion to the G0 star HD 106252. Fischer et al.

(2002) find that the dark companion has a minimum mass of 0.007 Msun, an orbital eccentricity of 0.57, and a semi-major axis of 2.42 AU. With these properties, the HD 106252 system appears to be more like a double star system than a star-planet system (Kumar 1964; Kumar 1974; Kumar 2002). In other words, HD 106252b is unlikely to have originated as a planet. The same conclusion may be drawn about many of the other dark companions that are being called 'extrasolar planets'.

## References

Fischer, D. A. et al. 2002, PASP, 114, 529
Kumar, S. S. 1962, AJ, 67, 579
Kumar, S. S. 1963a, ApJ, 137, 1121
Kumar, S. S. 1963b, ApJ, 137, 1126
Kumar, S. S. 1964, ZfAp, 58, 248
Kumar, S. S. 1967, Icarus, 6, 136
Kumar, S. S. 1974, Origins of Life, 5, 491
Kumar, S. S. 1990, Comments on Astrophysics, 15, 55
Kumar, S. S. 2002, Paper presented at IAU211 Symposium at Waikoloa, Hawaii (this volume)(http://lanl.arxiv.org/abs/astro-ph/0208096)